\documentclass[onecolumn,superscriptaddress,preprintnumbers,amsmath,amssymb,prc,nofootinbib]{revtex4}

\usepackage{graphicx}
\usepackage{subfigure}
\usepackage{dcolumn}
\usepackage{bm}

\usepackage{color}

\begin{document}
\title{Extracting jet transport parameter $\hat{q}$ from a multiphase transport model}

\author{Feng-Chu Zhou}
\affiliation{School of Physics and Electronic Science, Guizhou Normal University, Guiyang 550001, China}

\author{Guo-Liang Ma}
\email[]{glma@fudan.edu.cn}
\affiliation{Key Laboratory of Nuclear Physics and Ion-beam Application (MOE), Institute of Modern Physics, Fudan University, Shanghai 200433, China}
\affiliation{Shanghai Institute of Applied Physics, Chinese Academy of Sciences, Shanghai 201800, China}
\author{Yu-Gang Ma}
\email[]{mayugang@fudan.edu.cn}
\affiliation{Key Laboratory of Nuclear Physics and Ion-beam Application (MOE), Institute of Modern Physics, Fudan University, Shanghai 200433, China}
\affiliation{Shanghai Institute of Applied Physics, Chinese Academy of Sciences, Shanghai 201800, China}


\begin{abstract}

Within a multi-phase transport model with string melting scenario, jet transport parameter $\hat{q}$ is calculated in Au+Au collisions at $\sqrt{s_{NN} } $= 200 GeV  and Pb+Pb collisions at $\sqrt{s_{NN} } $= 2.76 TeV.  The $\hat{q}$ increases with the increasing of jet energy for both partonic phase and hadronic phase. The energy and path length dependences of $\hat{q}$ in full heavy-ion evolution are consistent with the expectations of jet quenching. The correlation between jet transport parameter  $\hat{q}$ and dijet transverse momentum asymmetry $A_{J}$ is mainly investigated, which discloses that a larger $\hat{q}$ corresponds to a larger $A_{J}$. It supports a consistent jet energy loss picture from the two viewpoints of single jet and dijet. It is proposed to measure dijet asymmetry distributions with different jet transport parameter ranges as a new potential method to study jet quenching physics in high energy heavy-ion collisions.

\end{abstract}


\maketitle

\section{Introduction}
\label{introduction}

The main goal of high energy heavy-ion collisions is to create a deconfined quark-gluon plasma (QGP) under an extreme temperature and(or) density condition~\cite{Collins:1974ky,Shuryak:1980tp,Luo:2017faz}. Jet is an important probe to search for the formed QGP  and investigate its properties,  because jets are originally generated from parton-parton hard interactions at the very early stage of heavy-ion collisions, while propagating through the expanding QGP medium. These hard jets lose energy through medium-induced gluon radiations and collisional energy loss,  known as jet quenching~\cite{Baier:1996sk,Blaizot:1986ma,Wang:1991xy,Gyulassy:1990ye,Baier:1994bd,Qin:2015srf}. The first evidence of the jet quenching phenomenon have been observed as the disappearance of away-side peak in dihadron correlation in central Au+Au collisions at the Relativistic Heavy Ion Collider(RHIC)~\cite{Adler:2002tq}.  The RHIC experiment has observed not only the suppression of single inclusive hadron spectrum at large transverse momentum $p_T$~\cite{Adams:2003kv,Adler:2003qi}, but also from back-to-back high-$p_{T}$ dihadron~\cite{Adler:2002tq} and $\gamma$-hadron correlation~\cite{Adare:2009vd,Abelev:2009gu}. The jet quenching phenomena are also widely studied in heavy-ion collisions at the Large Hadron Collider (LHC)~\cite{CMS:2012aa,Aamodt:2011vg}.  Recently, more and more results based on reconstructed jet measurements have been released, e.g., the suppression of reconstructed jets~\cite{Aad:2012vca} and $\gamma$-jet asymmetry~\cite{Chatrchyan:2012gt}.  A large dijet transverse momentum asymmetry has been observed by both ATLAS and CMS experiments at LHC~\cite{Aad:2010bu,Chatrchyan:2011sx}.  All kinds of observables  are consistent with the picture that jet significantly loses energy in the QGP. The more detailed studies of the propagation of jets inside the QGP could give us the more subtle aspects of jet quenching such as the parton mean free paths in the QGP and  the properties of the QGP itself.

Many theoritical models that incorporate parton energy loss have been proposed to study the observed jet quenching phenomena, such as: GLV~\cite{Gyulassy:2000er} and its CUJET implementation~\cite{Buzzatti:2011vt}, high twist approach (HT-BW and HT-M)~\cite{Guo:2000nz,Chen:2011vt,Majumder:2009ge,Vitev:2009rd}, MARTINI model~\cite{Schenke:2009gb}, McGill-AMY model~\cite{Qin:2007rn}, BAMPS model~\cite{Fochler:2010wn}, and LBT model~\cite{He:2015pra}.  Most of the theoretical models assumed a static potential for jet-medium interactions which result in a factorized dependence of parton energy loss on the jet transport parameter ($\hat{q}$).  The jet transport parameter $\hat{q}$, $i.e$, the mean-squared transverse momentum broadening per unit length,  is a widely used parameter that modulates the energy loss of  jet in a strong-interacting QCD medium,  which is also related to the distribution density of the medium and therefore characterizes the properties of the medium~\cite{Baier:1996sk,CasalderreySolana:2007sw}. Thus to investigate on jet energy dependence of transport parameter will not only improve our understanding of experimental results on jet quenching but also  can directly provide some information about the internal structure of the dense QCD matter by relating this parameter to the properties of the matter itself~\cite{CasalderreySolana:2007sw,Burke:2013yra}. On the other hand, One of the programmatic goals of heavy-ion collisions is to extract important medium properties from phenomenological studies of experimental data. Recent studies revel that low energy limit of the transport parameter of jet quenching is directly related to the shear viscosity ($\eta$/s) of the QGP~\cite{Majumder:2007zh,Liu:2006ug,Xu:2013yn}.  Since both $\eta$/s and $\hat{q}$ are transport parameters describing the exchange of energy and momentum between fast partons and medium, a pertinent question is whether there is a quantitative relation between these parameters that can be extracted from the hydrodynamic model and(or) dynamical transport model~\cite{Ayala:2016pvm}.  In short, systematic model studies on jet transport parameter are worth being further carried out.

The aim of this work is to give a detailed study of the jet transport parameter within A Multi-Phase Transport model (AMPT) with  string melting scenario.  The  study is carried out in two different collision systems at two different collision energies, $i.e.$ Au+Au collisions at $\sqrt{s_{NN} } $= 200 GeV  and Pb+Pb collisions at $\sqrt{s_{NN} } $= 2.76 TeV.  Since different collision energies can generate different system conditions such as different  temperatures in the overlap region, the results can provide some information about temperature dependence of jet observables. In order to understand the influence of environment effect on jet transport properties, this study is started from investigating the energy dependence of jet transport parameter, and then the path length dependence of the jet transport parameter is also investigated. Furthermore  the relationship between jet transport parameter, $\hat{q}$, and dijet transverse momentum asymmetry ,$A_{J}$, is carried out to understand and testify the consistence of jet quenching picture,  because dijet transverse momentum asymmetry is an important evidence of jet quenching pattern from the view of dijet while jet transport parameter is usually presented in the point of view of single jet. Because we find that the distributions of dijet transverse momentum asymmetry are different for different jet transport parameter ranges,  we propose a new method to study jet transport parameter $\hat{q}$ based on the dijet transverse momentum asymmetry $A_{J}$ distribution.  

This paper is organized as follows. We give a general setup of our modeling and calculations in Sec. \ref{GS}. In Sec. \ref{results}, we present our numerical results. We expand discussions and summarize in Sec. \ref{summary}.
\section{GENERAL SETUP}
\label{GS}
\subsection{AMPT model with a jet triggering technique}
The AMPT model with a string-melting mechanism as a dynamical transport model~\cite{Lin:2004en},  which has well described many experimental results~\cite{Zhang:2005ni,Ma:2011uma,Ma:2010dv}, is employed in this work. The model consists of four main phase gradients: initial condition, parton cascade, hadronization, and hadronic rescatterings. The initial condition, which includes the spatial and momentum distributions of participant matter, including minijet partons and soft string excitations, is obtained through the HIJING model~\cite{Wang:1991hta,Gyulassy:1994ew}. The parton cascade starts the partonic evolution with a quark-antiquark plasma from the melting of strings. Parton cascade is modeled by Zhang's parton cascade (ZPC)~\cite{Zhang:1997ej}, where the value of strong coupling constant and the Debye screening mass determine parton interaction cross section.  Only elastic collisions are included into the process of parton cascade at present.  A  quark coalescence model is then used to combine partons into hadrons~\cite{Lin:2001zk} when the partonic system freezes out. The evolution dynamics of the hadronic matter is described by a relativistic transport (ART) model~\cite{Li:1995pra}.

Since the QCD cross section of dijet production in AMPT model is quite small especially for high transverse momentum,  a jet trigger technique in HIJING model is used for our purpose of studying jet quenching physics.  Several hard dijet production channels are taken into account in the HIJING part of AMPT model including $q_{1} + q_{2}  \rightarrow q_{1} + q_{2}$,  $q_{1} + \overline{q_{1}}  \rightarrow q_{2} + \overline{q_{2}}$, $q + \overline{q}   \rightarrow g + g$, $q + g   \rightarrow q + g$, $g + g  \rightarrow q + \overline{q}$, and $g + g  \rightarrow g + g$~\cite{Sjostrand:1993yb}. The high transverse momentum primary partons pullulate to form jet parton showers through initial and final state QCD radiations. The jet parton showers are converted into clusters of on-shell quarks and anti-quarks through the string-melting mechanism of AMPT model. After the melting process, not only a quark and anti-quark plasma is formed , but also jet quark showers are built up, therefore the initial configuration between jets and medium is ready to interact.  All possible elastic partonic interactions among medium parton, between jet shower partons and medium partons, and among jet shower itself, are automatically simulated by the ZPC model.  After all partonic scatterings, the partons are recombined into medium hadrons or jet shower hadrons including the recombinations among shower partons and between shower partons and medium partons. In the hadronic phase, jet hadron shower continues to interact with the hadronic medium, which is described by ART model. The jet-triggered AMPT model has successfully given qualitative descriptions to many experimental results on reconstructed jet observables~\cite{Nie:2014nst}, such as $\gamma$-jet imbalance~\cite{Ma:2013bia}, dijet asmmetry~\cite{Ma:2013pha}, decomposition of jet fragmentation function~\cite{Ma:2013gga,Ma:2013yoa} and jet shape~\cite{Ma:2013uqa}, by including jet collisional energy loss only with a large interaction cross section (1.5 mb). However, it is still questionable whether the AMPT model can give a reasonable jet transport parameter $\hat{q}$. Therefore, the main task of this work is to extract the jet transport parameter $\hat{q}$ and investigate its properties in high energy heavy-ion collisions, within the framework of the AMPT model.

\subsection{Calculations of the  transport parameter $\hat{q}$}

The definition of jet transport parameter, $\hat{q}$, is the average transverse momentum squared per unit length transferred  from jet to the medium which reads as follows:
\begin{eqnarray}
\hat{q} = \frac{<p_{\perp}^{2}>}{L},
\label{qhat_definition}
\end{eqnarray}%
where $p_{\perp} =  p_{jet} sin\theta$ is the jet transverse momentum broadening due to the interactions between jet and medium,  $p_{jet}$ is the final jet momentum after jet-medium interactions, and $\theta$ is the relative angle of identical jet momentum between  before and after jet-medium interactions, and $L$ is the path length that jet travels through the medium.

In this work, our method of extracting $\hat{q}$  is designed based on the information about the time evolution of jet-medium interactions in AMPT model.  For simplicity, we only  illustrate how to apply it for extracting $\hat{q}$ in partonic phase.  The first step is to reconstructed jets based on the information about the partons before parton cascade, where we use the anti-$k_{t}$ algorithm from the standard Fastjet package for jet reconstruction~\cite{Cacciari:2011ma,hep-ph/0512210}.  These reconstructed jets at this stage has not yet undergone any interactions, thus carry initial state information.  The second step is to reconstruct jets from the partons after parton cascade, which find  jets that have gone through the whole partonic  evolution. The third step is to pairing the initial jets and final jets. We require that the pair with a relative pseudorapidity gap ($\Delta\eta$ ) and relative azimuthal angle  ($\Delta\phi$) fulfill the selection of  $|\Delta\eta| < 0.2$  and  $|\Delta\theta| < 0.2$. The good pairing guarantees that two jets in one pair are an identical same jet but reconstructed at two different stages for a same event.  All gradients for extracting $\hat{q}$ are now ready except the jet path length in coordinate space. The jet coordinate is represented by its leading particle whose energy is required to carry at least one third of the whole jet energy in our analysis. The path length is obtained by calculating the distance between two vertexes of jets in pair in coordinate space.

To acquire  qualified jet for this study, the kinetic cut for jet reconstruction are choose as the general requirement as in experiments. The jet cone(R) is set to 0.5. The  transverse momentum of jet is required to be larger than 40 GeV/c ( $p_{T} > 120 GeV/c$ in the case of $A_{J}$  calculation) for leading jet, while $p_{T} > 20 GeV/c$ ( $p_{T} > 50 GeV/c$ in the case of $A_{J}$  calculation) for subleading jet. And the pseudorapidity range of jet within $|\eta| < 2$ is required in our analysis.  

\section{Results and discussions}
\label{results}
\subsection{energy and path length dependences of jet transport parameter $\hat{q}$}
\label{resultsA}

\begin{figure}[htbp]
\begin{minipage}[t]{0.46\linewidth}
\centering
\includegraphics[width=1.04\textwidth]{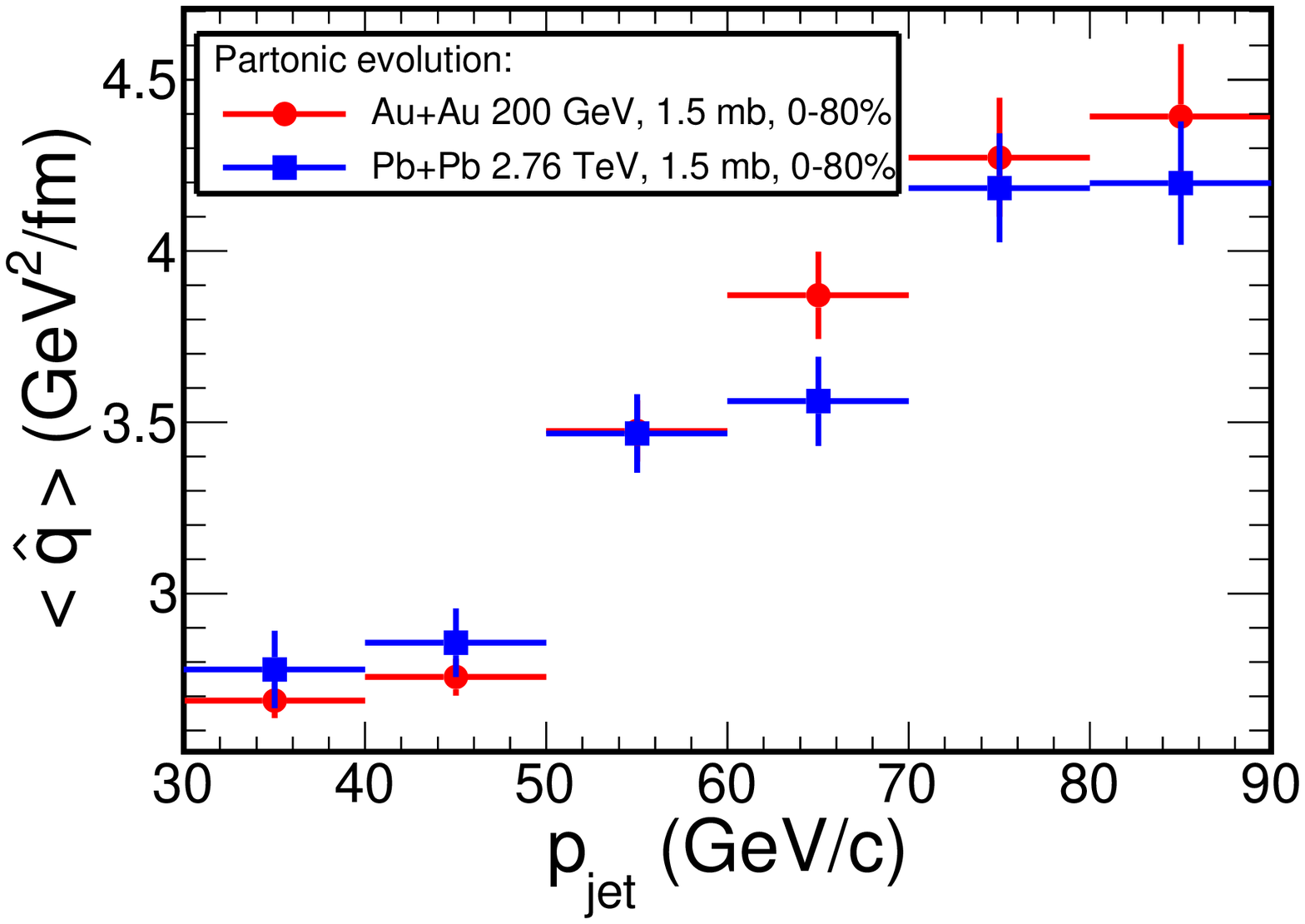}
\caption{The energy dependences of jet transport parameter  $\hat{q}$ in the partonic phase for Au+Au collisions at $\sqrt {s_{NN}} = 200$ GeV and Pb+Pb collisions at $\sqrt {s_{NN}} = 2.76$ TeV.}\label{fig:qhat_p:a}
\end{minipage}%
\hfill
\begin{minipage}[t]{0.46\linewidth}
\centering
\includegraphics[width=1.04\textwidth]{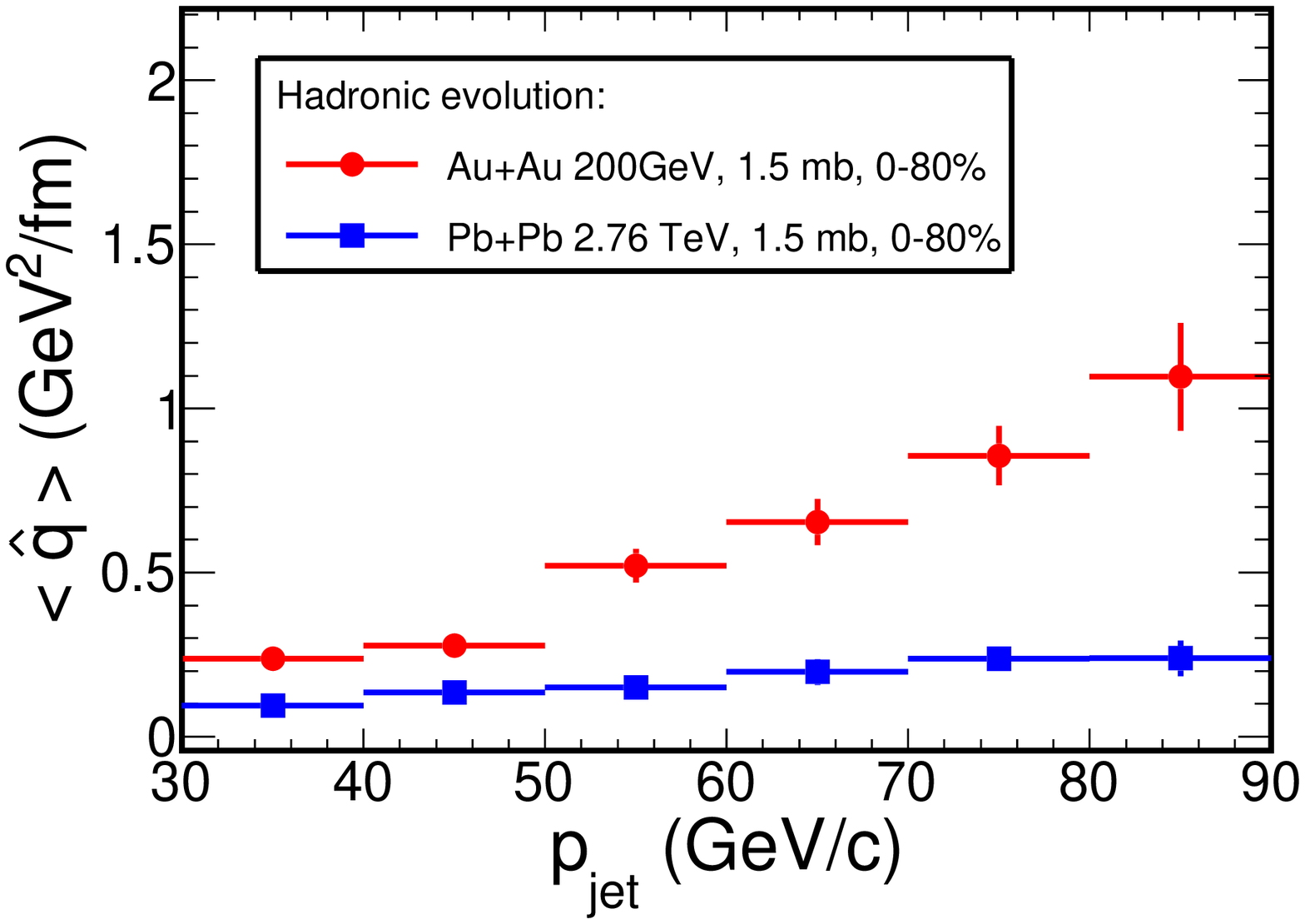}
\caption{ The energy dependences of jet transport parameter  $\hat{q}$  in the hadronic phase for Au+Au collisions at $\sqrt {s_{NN}} = 200$ GeV and Pb+Pb collisions at $\sqrt {s_{NN}} = 2.76$ TeV.}\label{fig:qhat_p:b}
\end{minipage}
\label{fig:qhat_p}
\end{figure}

Fig.~\ref{fig:qhat_p:a} shows the energy dependences of jet transport parameter $\hat{q}$ in the partonic phase for Au+Au collisions at $\sqrt s = 200$ GeV and Pb+Pb collisions at $\sqrt s = 2.76$ TeV.  We can see a significant increasing trend of $\hat{q}$ with the increasing of jet energy for  two different collision energies. It is consistent with the jet quenching picture that  higher energy jets have   larger transverse momentum transfer from jet to medium~\cite{CasalderreySolana:2007sw}. We also notice that the  $\hat{q}$ extracted from  $\sqrt s = 2.76$ TeV  Pb+Pb collision is slightly larger than that from $\sqrt s = 200$ GeV  Au+Au collisions at low  jet energy while close to each other at high jet energy. It indicates that low-energy jets are more sensitive to the properties (like temperature)  of the medium than high-energy jets.

Fig.~\ref{fig:qhat_p:b} shows that $\hat{q}$ in hadronic phase also increases with the increasing of jet energy for two collisions energies. It indicates that jets continue to interacts with hadronic medium which can also induces jet transverse momentum broadening.  We see that the magnitudes of the $\hat{q}$  extracted from hadronic phase are less than those extracted from  partonic phase. The difference in magnitude reveals that jets undergo stronger interactions in partonic phase, since the partonic medium is much hotter and denser than the hadronic medium. It indicates the partonic interactions dominate the jet transverse momentum broadening. We also notice that the hadronic $\hat{q}$  from Au+Au collisions at $\sqrt s = 200$ GeV is larger than that for Pb+Pb collisions at $\sqrt s = 2.76$ TeV,  which may be because the system becomes more diluted after a long-lived expanding partonic phase in Pb+Pb collisions than that in Au+Au collisions.

\begin{figure}[htbp]
\begin{minipage}[t]{0.46\linewidth}
\centering
\includegraphics[width=1.04\textwidth]{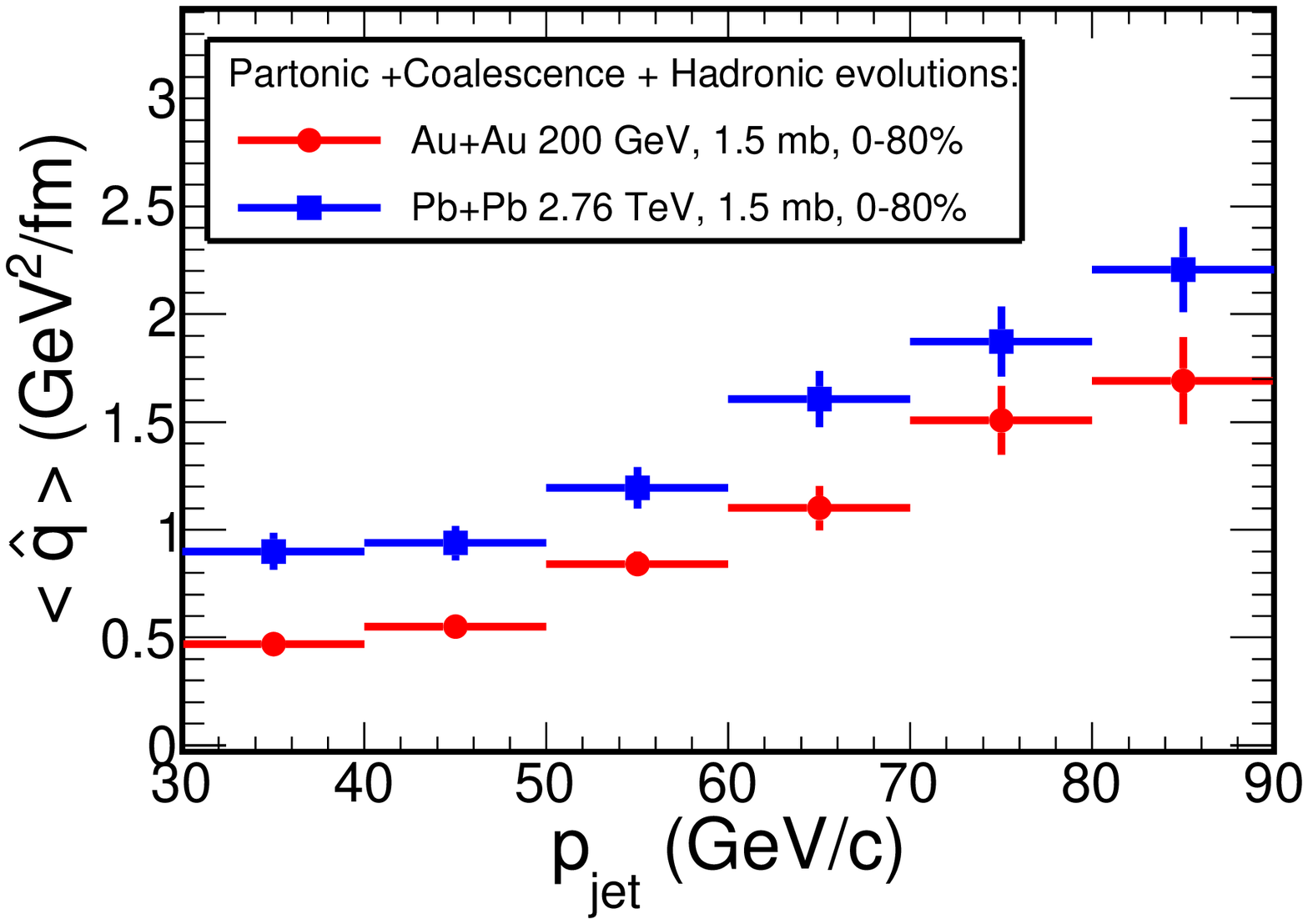}
\caption{The energy dependences of transport parameter  $\hat{q}$  in the full evolution for Au+Au collisions at $\sqrt {s_{NN}} = 200$ GeV and Pb+Pb collisions at $\sqrt {s_{NN}} = 2.76$ TeV.}\label{fig:qhat_p}
\end{minipage}%
\hfill
\begin{minipage}[t]{0.46\linewidth}
\centering
\includegraphics[width=1.04\textwidth]{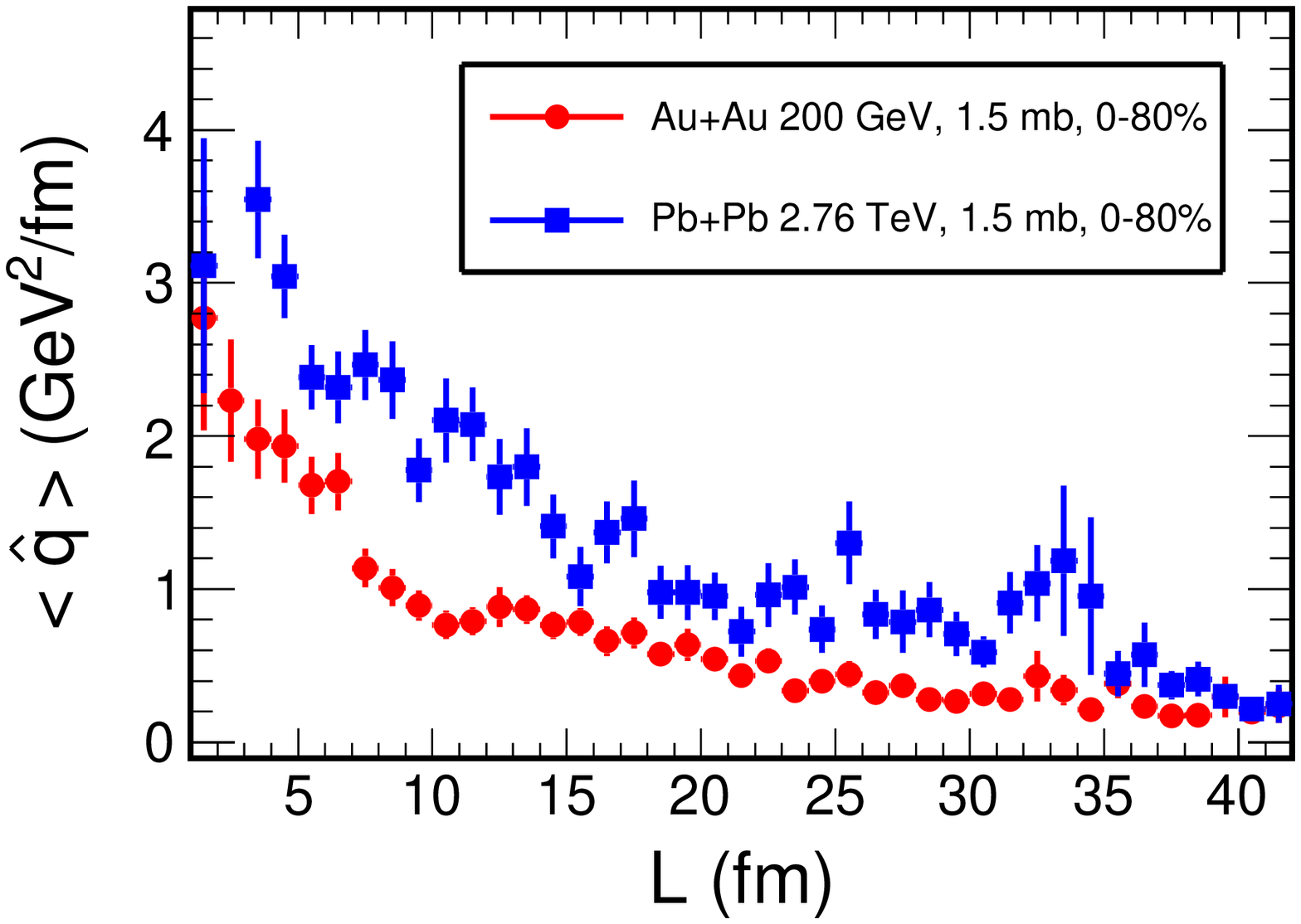}
\caption{ The travel path length dependences of transport parameter  $\hat{q}$ in the full evolution for Au+Au collisions at $\sqrt {s_{NN}} = 200$ GeV and Pb+Pb collisions at $\sqrt {s_{NN}} = 2.76$ TeV.}\label{fig:qhat_L}
\end{minipage}
\end{figure}

We also extract the $\hat{q}$ from the full evolution of heavy-ion collisions by using the information about identical jets before parton cascade and after hadronic phase evolution. In Fig.~\ref{fig:qhat_p},  a significant increasing trend of $\hat{q}$ with the increasing of jet energy for two different collision energies is also observed in the full evolution. We observe a hierarchy of $\hat{q}$ in the full evolution, which shows that a higher collision energy has a larger $\hat{q}$.  In Fig.~\ref{fig:qhat_p},  we consider the transverse momentum broadening and travel length in the full evolution instead of separating them to two pieces for partonic and hadronic phases. We should emphasize that the $\hat{q}$ extracted from the full evolution is not a simple sum of the two $\hat{q}$ for partonic and hadronic phases. Note that the partonic phase evolution brings a  larger transverse momentum broadening while the hadronic phase evolution provides a longer travel path length.

Fig.~\ref{fig:qhat_L} shows that $\hat{q}$ decreases with the increasing of travel path length for the full evolution.  Because $\hat{q}$ is defined as per unit length broadening of transverse momentum as in Eq.(~\ref{qhat_definition}), our results are consistent with the expectation.

\subsection{The jet transport parameter $\hat{q}$ and dijet  asymmetry $A_{J}$}
\label{resultsB}

\begin{figure*}[htb]
 \setlength{\abovecaptionskip}{0pt}
 \setlength{\belowcaptionskip}{8pt}\centerline{\includegraphics[scale=0.765]{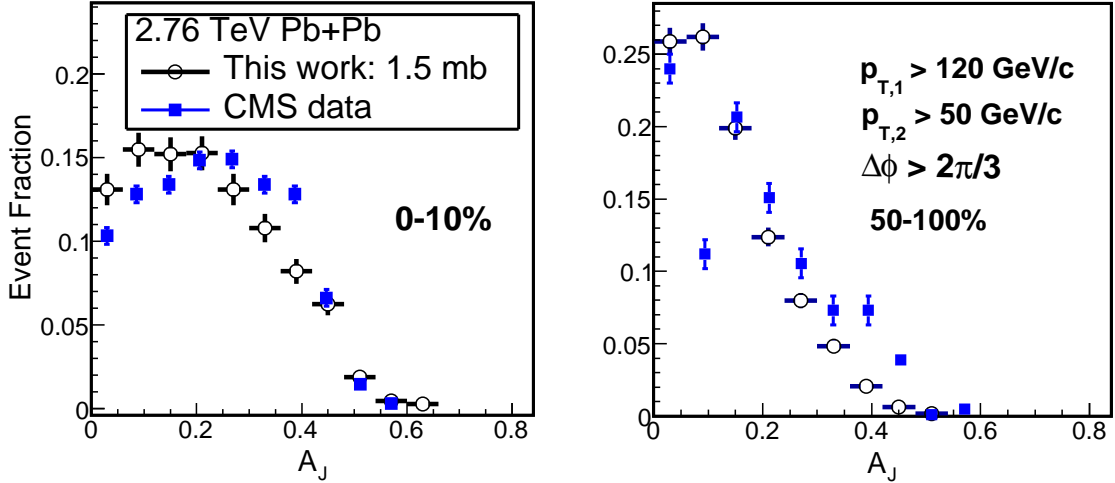}}
\caption{(Color online) The dijet asymmetry parameter $A_{J}$ distributions for the centrality bins of 0-10\% (left) and 50-100\% (right) in Pb+Pb collisions at $\sqrt s = 2.76$ TeV. The data points are taken from the CMS collaboration~\cite{Chatrchyan:2011sx}}\label{fig:Aj_reproduce}
\end{figure*}

There have been many experimental observables that disclose the natures of jet quenching from different points of view. Dijet transverse momentum asymmetry  $A_{J}$  is an important probe of jet quenching from the view of dijet picture where one selects the leading jet as a trigger and then looking into whether the  associated away-side jet suffers energy loss or not. The dijet asymmetry parameter $A_{J}$ is defined as follows,  
\begin{eqnarray}
A_{J} = \frac{p_{T,1} - p_{T,2}}{p_{T,1} + p_{T,2}} ,
\label{AJ_definition}
\end{eqnarray}%
where $p_{T,1} $ is the transverse momentum of the  leading jet while $p_{T,2}$ is transverse momentum of associated jet. The asymmetry parameter $A_{J}$ should be close to zero if there is no jet energy loss, because the transverse momentum of leading jet and associated jet should be close to each other if there is no medium effect  in a system. While if the associated jet travels through medium with a longer path length hence suffers more energy loss, its transverse momentum will significantly be reduced relative to the transverse momentum of leading jet  thus $A_{J}$ will differ from zero. In this study, we first reproduce the results published by the CMS collaboration in Ref.~\cite{Chatrchyan:2011sx} to make sure our method relaible. As shown in the Fig.~\ref{fig:Aj_reproduce},  our results can well reproduce the CMS data not only for central events ($0-10\%$) but also for peripheral events ($50-100\%$) in Pb+Pb 2.76 TeV collisions~\cite{Ma:2013pha}.

\begin{figure}[htbp]
\begin{minipage}[t]{0.46\linewidth}
\centering
\includegraphics[width=1.04\textwidth]{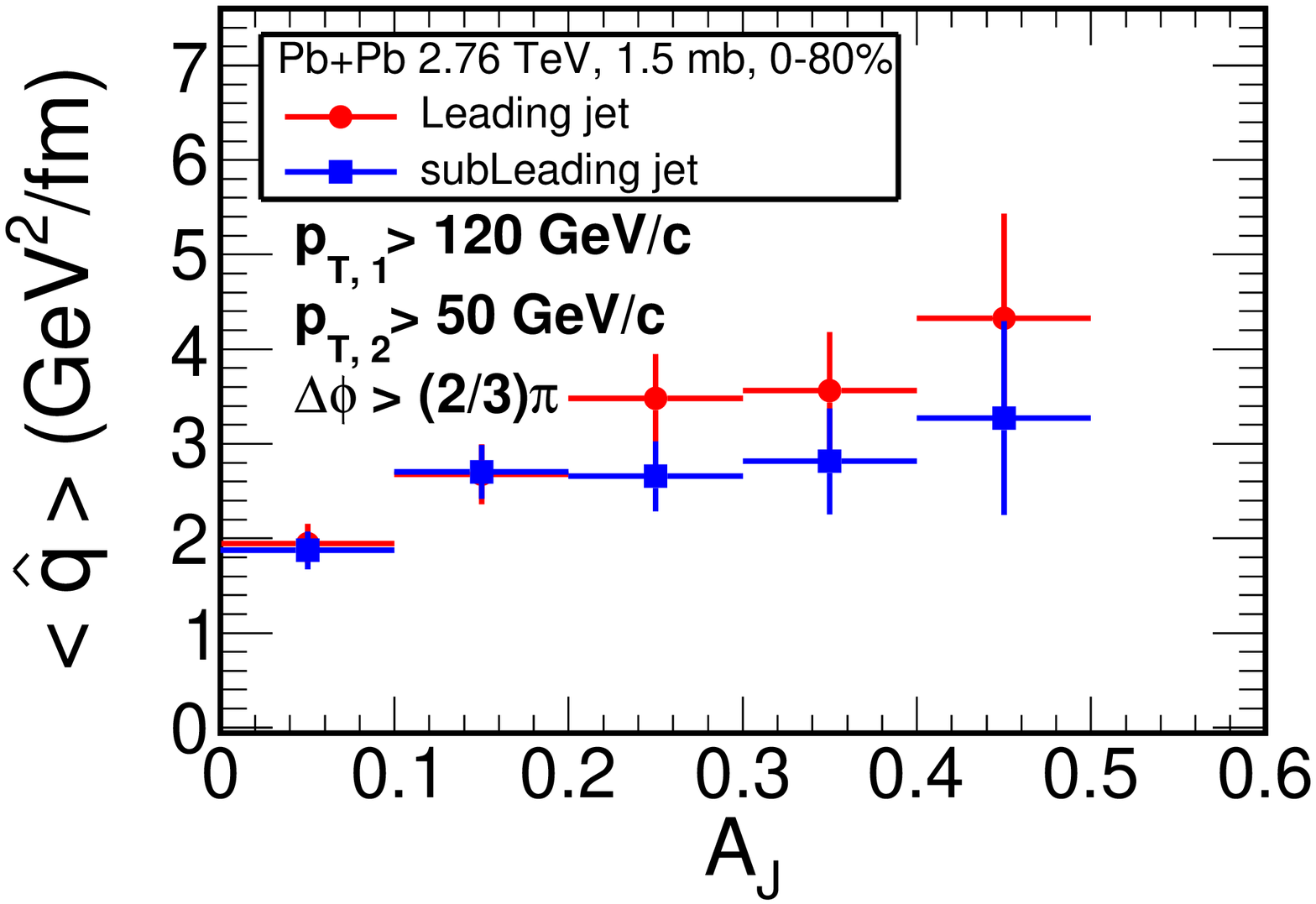}
\caption{(Color online) The jet transport parameter as function of dijet asymmetry parameter for Pb+Pb collisions at $\sqrt s = 2.76$ TeV.}\label{fig:Aj_qhat}
\end{minipage}%
\hfill
\begin{minipage}[t]{0.46\linewidth}
\centering
\includegraphics[width=1.04\textwidth]{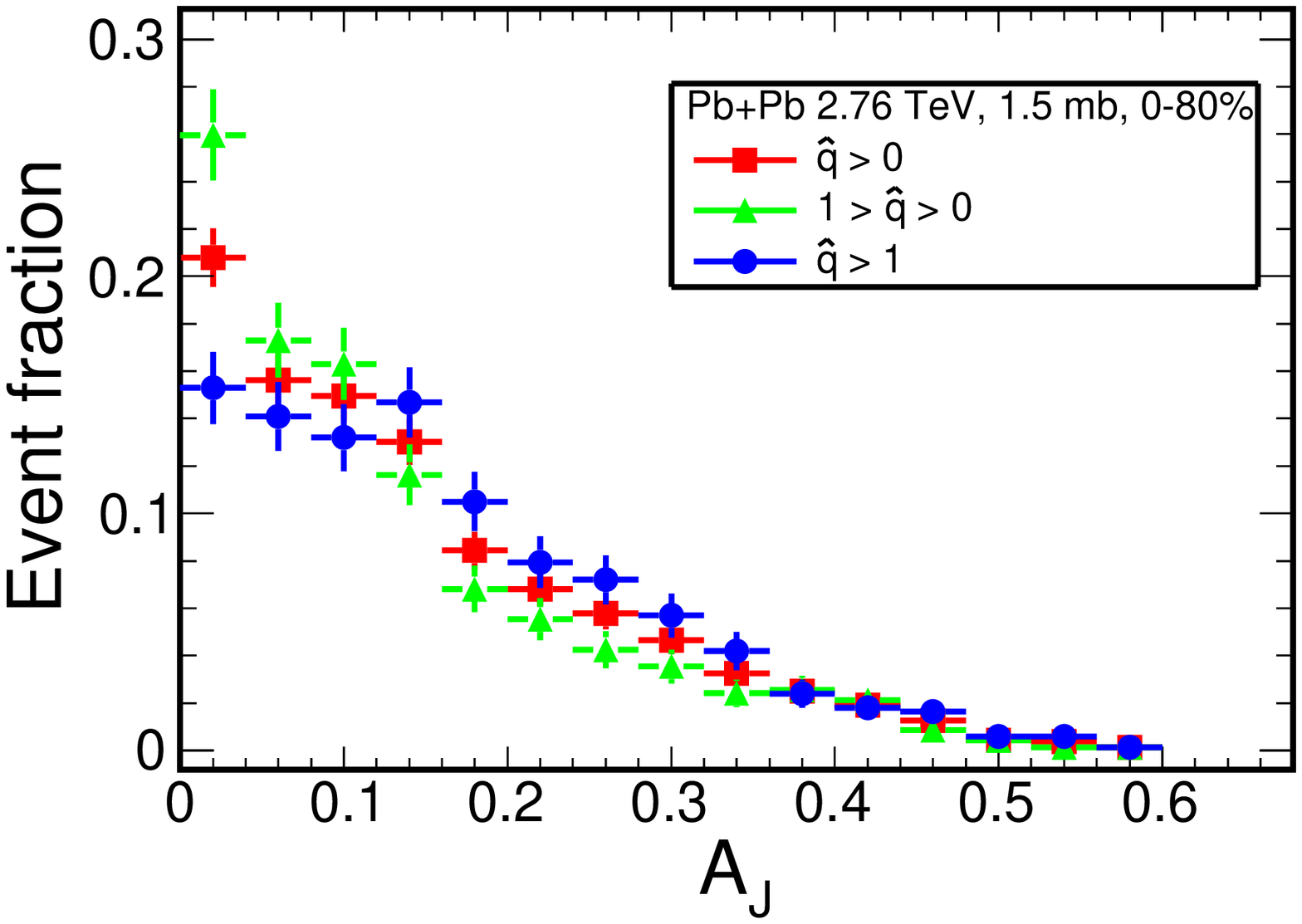}
\caption{(Color online) The dijet asymmetry parameter $A_J$ distributions for different jet transport parameter ranges for Pb+Pb collisions at $\sqrt s = 2.76$ TeV.}\label{fig:Aj_qhat_diff}
\end{minipage}%
\end{figure}

In order to investigate the $\hat{q}$  from the point of view  of dijet asymmetry,  the correlation between jet transport parameter  $\hat{q}$  and dijet transverse momentum  $A_{J}$ is studied.  Fig.~\ref{fig:Aj_qhat}  shows that the $\hat{q}$ increases with the increasing of the $A_{J}$ for both leading jet and subleading jet, which is consistent with the jet quenching picture that a larger jet energy loss gives a larger $\hat{q}$. In Fig.~\ref{fig:Aj_qhat}, we also noticed that $\hat{q}$ for leading and subleading jets get converged when $A_{J}$ closed to zero,  while get separated when the $A_{J}$ increases. It can be understood since a lower value of $A_{J}$ means a less transverse momentum asymmetry between leading and subleading jets which actually indicates that two jets are influenced by the medium with a same amount of magnitude, thus  they should have a similar $\hat{q}$. However, the increasing of $A_{J}$ means the larger and larger transverse momentum asymmetry due to the different magnitudes of jet energy loss that leading and subleading jets suffer, thus this difference of their energy loss magnitude makes the  $\hat{q}$ of leading jet and subleading jet get separated in the large $A_{J}$ range. There is also an interesting feature in Fig.~\ref{fig:Aj_qhat}  that  the magnitude of $\hat{q}$ on leading jet case is always larger than subleading jet case. The reason is because the leading jet generally hold a larger energy than subleading jet.  From Fig.~\ref{fig:qhat_p}, we know that the $\hat{q}$  increases with the increasing of jet energy thus the magnitude of $\hat{q}$ is always larger for leading jet which carries a larger energy than sub-leading jet. The consistence between jet transport parameter  $\hat{q}$ and dijet transverse momentum  $A_{J}$ from two different points of view of single jet and dijet is a strong evidence that supports the physical picture of jet quenching in the hot and dense QCD matter .

The jet transport parameter $\hat{q}$ is such an important parameter that both theoretical and experimental communities are greatly interested on that. In many theoretical models, $\hat{q}$ is somehow  treated as a model parameter that control the degree of jet quenching, where one can tune the $\hat{q}$ parameter to calculate some observables such as nuclear modification factor ($R_{AA}$)  to match the experimental data hence iteratively back to decide a more proper value of $\hat{q}$~\cite{Burke:2013yra}.  Recently, there are also some $\hat{q}$ extracting measurements by using di-hadron correlations and hadron-jet correlations~\cite{Tannenbaum:2017afg,Chen:2016vem,Mueller:2016gko}. The correlation between jet transport parameter $<\hat{q}>$  and dijet transverse momentum asymmetry $A_{J}$  may stimulates ones to raise a new method of constraining and even extracting the jet transport parameter $<\hat{q}>$ from the dijet asymmetry measurements. In Fig.~\ref{fig:Aj_qhat_diff}, we show the $A_{J}$ distributions for several different $\hat{q}$ ranges. It is consistent with the jet quenching picture that the $A_{J}$ distribution with a larger $\hat{q}$ is shifted right to a higher $A_{J}$ value rather than that with a smaller $\hat{q}$. Therefore, it is very potential to extract the jet transport parameter $<\hat{q}>$ (and even some properties of the QGP, e.g. $\eta$/s) form the comparison of theoretical jet-quenching models and the experimental data based on dijet asymmetry $A_{J}$  distribution.

\section{Conclusions}
\label{summary}
In summary, we have utilized the AMPT model to investigate the jet energy dependence of jet transport parameter $\hat{q}$ in Au+Au collisions at $\sqrt{s_{NN} } $= 200 GeV  and Pb+Pb collisions at $\sqrt{s_{NN} } $= 2.76 TeV.  The  significant increase of $\hat{q}$ with the increasing of jet energy in partonic phase evolution while a less significant dependence in hadronic phase are observed. The same behavior is also observed in the full evolution of heavy-ion collisions. Furthermore, the correlation between jet transport parameter $\hat{q}$ and dijet transverse momentum asymmetry  $A_{J}$ is investigated in order to understand the jet quenching picture from two different points of view.  The increase of  $\hat{q}$  with increasing of $A_{J}$ is observed for leading and subleading jets, indicating the consistence between the perspectives of single jet and dijet.  Finally,  the dijet transverse momentum asymmetry distributions with different jet transport parameter $\hat{q}$ ranges show a clear deviation. This inspiredly leads to a new proposed method of constraining or extracting jet transport parameter $\hat{q}$ from experiment data by measuring the $A_{J}$ distribution in a comparison with theoretical calculations.

\begin{acknowledgments}
F.-C.Z. is grateful to the Shanghai Institute of Applied Physics for its hospitality. F.-C.Z. is supported by the National Natural Science Foundation of China under Grants 11447203, the Science and Technology Department of Guizhou Province Fund under Grant Nos [2014]7053, and the Doctoral Research Fund of Guizhou Normal University. G.-L.M. and Y.-G.M. are supported by the National Natural Science Foundation of China under Grants No. 11890714, 11421505, 11835002, 11522547 and 11375251, the Key Research Program of the Chinese Academy of Sciences under Grant No. XDPB09.

\end{acknowledgments}


\begin{thebibliography}{99}

\bibitem{Collins:1974ky} 
  J.~C.~Collins and M.~J.~Perry,
  Phys.\ Rev.\ Lett.\  {\bf 34}, 1353 (1975).
  
\bibitem{Shuryak:1980tp} 
  E.~V.~Shuryak,
  Phys.\ Rept.\  {\bf 61}, 71 (1980).
  
\bibitem{Luo:2017faz} 
  X.~Luo and N.~Xu,
  Nucl.\ Sci.\ Tech.\  {\bf 28}, no. 8, 112 (2017)
  [arXiv:1701.02105 [nucl-ex]].

\bibitem{Baier:1996sk} 
  R.~Baier, Y.~L.~Dokshitzer, A.~H.~Mueller, S.~Peigne and D.~Schiff,
  Nucl.\ Phys.\ B {\bf 484}, 265 (1997)
  [hep-ph/9608322].

\bibitem{Blaizot:1986ma}
  J.~P.~Blaizot and L.~D.~McLerran,
  Phys.\ Rev.\ D {\bf 34} (1986) 2739.

\bibitem{Wang:1991xy} 
  X.~N.~Wang and M.~Gyulassy,
  Phys.\ Rev.\ Lett.\  {\bf 68}, 1480 (1992).

\bibitem{Gyulassy:1990ye} 
  M.~Gyulassy and M.~Plumer,
  Phys.\ Lett.\ B {\bf 243}, 432 (1990).

\bibitem{Baier:1994bd} 
  R.~Baier, Y.~L.~Dokshitzer, S.~Peigne and D.~Schiff,
  Phys.\ Lett.\ B {\bf 345}, 277 (1995)
  [hep-ph/9411409].

\bibitem{Qin:2015srf} 
  G.~Y.~Qin and X.~N.~Wang,
  Int.\ J.\ Mod.\ Phys.\ E {\bf 24}, no. 11, 1530014 (2015)
  [arXiv:1511.00790 [hep-ph]].
  
\bibitem{Adler:2002tq} 
  C.~Adler {\it et al.} [STAR Collaboration],
  Phys.\ Rev.\ Lett.\  {\bf 90}, 082302 (2003)
  [nucl-ex/0210033].

\bibitem{Adams:2003kv} 
  J.~Adams {\it et al.} [STAR Collaboration],
  Phys.\ Rev.\ Lett.\  {\bf 91}, 172302 (2003)
  [nucl-ex/0305015].

\bibitem{Adler:2003qi} 
  S.~S.~Adler {\it et al.} [PHENIX Collaboration],
  Phys.\ Rev.\ Lett.\  {\bf 91}, 072301 (2003)
  [nucl-ex/0304022].

\bibitem{Adare:2009vd} 
  A.~Adare {\it et al.} [PHENIX Collaboration],
  Phys.\ Rev.\ C {\bf 80}, 024908 (2009)
  [arXiv:0903.3399 [nucl-ex]].

\bibitem{Abelev:2009gu} 
  B.~I.~Abelev {\it et al.} [STAR Collaboration],
  Phys.\ Rev.\ C {\bf 82}, 034909 (2010)
  [arXiv:0912.1871 [nucl-ex]].

\bibitem{CMS:2012aa} 
  S.~Chatrchyan {\it et al.} [CMS Collaboration],
  Eur.\ Phys.\ J.\ C {\bf 72}, 1945 (2012)
  [arXiv:1202.2554 [nucl-ex]].

\bibitem{Aamodt:2011vg} 
  K.~Aamodt {\it et al.} [ALICE Collaboration],
  Phys.\ Rev.\ Lett.\  {\bf 108}, 092301 (2012)
  [arXiv:1110.0121 [nucl-ex]].
  
\bibitem{Aad:2012vca} 
  G.~Aad {\it et al.} [ATLAS Collaboration],
  Phys.\ Lett.\ B {\bf 719}, 220 (2013)
  [arXiv:1208.1967 [hep-ex]].
  
\bibitem{Chatrchyan:2012gt} 
  S.~Chatrchyan {\it et al.} [CMS Collaboration],
  Phys.\ Lett.\ B {\bf 718}, 773 (2013)
  [arXiv:1205.0206 [nucl-ex]].
  
\bibitem{Aad:2010bu} 
  G.~Aad {\it et al.} [ATLAS Collaboration],
  Phys.\ Rev.\ Lett.\  {\bf 105}, 252303 (2010)
  [arXiv:1011.6182 [hep-ex]].
  
\bibitem{Chatrchyan:2011sx} 
  S.~Chatrchyan {\it et al.} [CMS Collaboration],
  Phys.\ Rev.\ C {\bf 84}, 024906 (2011)
  [arXiv:1102.1957 [nucl-ex]].

\bibitem{Gyulassy:2000er} 
  M.~Gyulassy, P.~Levai and I.~Vitev,
  Nucl.\ Phys.\ B {\bf 594}, 371 (2001)
  [nucl-th/0006010].
  
\bibitem{Buzzatti:2011vt} 
  A.~Buzzatti and M.~Gyulassy,
  Phys.\ Rev.\ Lett.\  {\bf 108}, 022301 (2012)
  [arXiv:1106.3061 [hep-ph]].
  
\bibitem{Guo:2000nz} 
  X.~f.~Guo and X.~N.~Wang,
  Phys.\ Rev.\ Lett.\  {\bf 85}, 3591 (2000)
  [hep-ph/0005044].
  
\bibitem{Chen:2011vt} 
  X.~F.~Chen, T.~Hirano, E.~Wang, X.~N.~Wang and H.~Zhang,
  Phys.\ Rev.\ C {\bf 84}, 034902 (2011)
  [arXiv:1102.5614 [nucl-th]].
  
\bibitem{Majumder:2009ge} 
  A.~Majumder,
  Phys.\ Rev.\ D {\bf 85}, 014023 (2012)
  [arXiv:0912.2987 [nucl-th]].
  
\bibitem{Vitev:2009rd} 
  I.~Vitev and B.~W.~Zhang,
  Phys.\ Rev.\ Lett.\  {\bf 104}, 132001 (2010)
  [arXiv:0910.1090 [hep-ph]].
  
\bibitem{Schenke:2009gb} 
  B.~Schenke, C.~Gale and S.~Jeon,
  Phys.\ Rev.\ C {\bf 80}, 054913 (2009)
  [arXiv:0909.2037 [hep-ph]].
  
\bibitem{Qin:2007rn} 
  G.~Y.~Qin, J.~Ruppert, C.~Gale, S.~Jeon, G.~D.~Moore and M.~G.~Mustafa,
  Phys.\ Rev.\ Lett.\  {\bf 100}, 072301 (2008)
  [arXiv:0710.0605 [hep-ph]].

\bibitem{Fochler:2010wn} 
  O.~Fochler, Z.~Xu and C.~Greiner,
  Phys.\ Rev.\ C {\bf 82}, 024907 (2010)
  [arXiv:1003.4380 [hep-ph]].
  
\bibitem{He:2015pra} 
  Y.~He, T.~Luo, X.~N.~Wang and Y.~Zhu,
  Phys.\ Rev.\ C {\bf 91}, 054908 (2015)
  Erratum: [Phys.\ Rev.\ C {\bf 97}, no. 1, 019902 (2018)]
  [arXiv:1503.03313 [nucl-th]].
  
\bibitem{CasalderreySolana:2007sw} 
  J.~Casalderrey-Solana and X.~N.~Wang,
  Phys.\ Rev.\ C {\bf 77}, 024902 (2008)
  [arXiv:0705.1352 [hep-ph]].

\bibitem{Burke:2013yra} 
  K.~M.~Burke {\it et al.} [JET Collaboration],
  Phys.\ Rev.\ C {\bf 90}, no. 1, 014909 (2014)
  [arXiv:1312.5003 [nucl-th]].
  
\bibitem{Majumder:2007zh} 
  A.~Majumder, B.~Muller and X.~N.~Wang,
  Phys.\ Rev.\ Lett.\  {\bf 99}, 192301 (2007)
  [hep-ph/0703082].
  
\bibitem{Liu:2006ug} 
  H.~Liu, K.~Rajagopal and U.~A.~Wiedemann,
  Phys.\ Rev.\ Lett.\  {\bf 97}, 182301 (2006)
  [hep-ph/0605178].
  
\bibitem{Xu:2013yn} 
  J.~Xu,
  Nucl.\ Sci.\ Tech.\  {\bf 24}, no. 5, 50514 (2013)
  [arXiv:1302.0165 [nucl-th]].
  
\bibitem{Ayala:2016pvm} 
  A.~Ayala, I.~Dominguez, J.~Jalilian-Marian and M.~E.~Tejeda-Yeomans,
  Phys.\ Rev.\ C {\bf 94}, no. 2, 024913 (2016)
  [arXiv:1603.09296 [hep-ph]].
  
  \bibitem{Lin:2004en}
  Z.~W.~Lin, C.~M.~Ko, B.~A.~Li, B.~Zhang and S.~Pal,
  Phys.\ Rev.\ C {\bf 72}, 064901 (2005)
  [nucl-th/0411110].
  
\bibitem{Zhang:2005ni} 
  B.~Zhang, L.~W.~Chen and C.~M.~Ko,
  Phys.\ Rev.\ C {\bf 72}, 024906 (2005)
  [nucl-th/0502056].
  
\bibitem{Ma:2011uma} 
  G.~L.~Ma and B.~Zhang,
  Phys.\ Lett.\ B {\bf 700}, 39 (2011)
  [arXiv:1101.1701 [nucl-th]].
  
\bibitem{Ma:2010dv} 
  G.~L.~Ma and X.~N.~Wang,
  Phys.\ Rev.\ Lett.\  {\bf 106}, 162301 (2011)
  [arXiv:1011.5249 [nucl-th]].
  
   \bibitem{Wang:1991hta}
  X.~N.~Wang and M.~Gyulassy,
  Phys.\ Rev.\ D {\bf 44}, 3501 (1991).
  
\bibitem{Gyulassy:1994ew}
  M.~Gyulassy and X.~N.~Wang,
  Comput.\ Phys.\ Commun.\  {\bf 83}, 307 (1994)
  [nucl-th/9502021].
  
  \bibitem{Zhang:1997ej}
  B.~Zhang,
  Comput.\ Phys.\ Commun.\  {\bf 109}, 193 (1998)
  [nucl-th/9709009].
\bibitem{Lin:2001zk}
  Z.~W.~Lin and C.~M.~Ko,
  Phys.\ Rev.\ C {\bf 65}, 034904 (2002)
  [nucl-th/0108039].
\bibitem{Li:1995pra}
  B.~A.~Li and C.~M.~Ko,
  Phys.\ Rev.\ C {\bf 52}, 2037 (1995)
  [nucl-th/9505016].

\bibitem{Sjostrand:1993yb} 
  T.~Sjostrand,
  Comput.\ Phys.\ Commun.\  {\bf 82}, 74 (1994).
  doi:10.1016/0010-4655(94)90132-5
  
\bibitem{Nie:2014nst} 
  M.~W.~Nie and G.~L.~Ma,
  Nucl. Tech. {\bf 37}, 100519 (2014).
    
\bibitem{Ma:2013bia} 
  G.~L.~Ma,
  Phys.\ Lett.\ B {\bf 724}, 278 (2013)
  [arXiv:1302.5873 [nucl-th]].
  
\bibitem{Ma:2013pha} 
  G.~L.~Ma,
  Phys.\ Rev.\ C {\bf 87}, no. 6, 064901 (2013)
  [arXiv:1304.2841 [nucl-th]].
  
\bibitem{Ma:2013gga} 
  G.~L.~Ma,
  Phys.\ Rev.\ C {\bf 88}, no. 2, 021902 (2013)
  [arXiv:1306.1306 [nucl-th]].
  
\bibitem{Ma:2013yoa} 
  G.~L.~Ma,
  Phys.\ Rev.\ C {\bf 89}, no. 6, 064909 (2014)
  [arXiv:1310.3701 [nucl-th]].
  
\bibitem{Ma:2013uqa} 
  G.~L.~Ma,
  Phys.\ Rev.\ C {\bf 89}, no. 2, 024902 (2014)
  [arXiv:1309.5555 [nucl-th]].
  
\bibitem{Cacciari:2011ma} 
  M.~Cacciari, G.~P.~Salam and G.~Soyez,
  Eur.\ Phys.\ J.\ C {\bf 72}, 1896 (2012)
  [arXiv:1111.6097 [hep-ph]].
  
\bibitem{hep-ph/0512210}
  M.~Cacciari and G.~P.~Salam,
  Phys.\ Lett.\ B\ {\bf 641} (2006) 57
  [hep-ph/0512210].
  
\bibitem{Tannenbaum:2017afg} 
  M.~J.~Tannenbaum,
  Phys.\ Lett.\ B {\bf 771}, 553 (2017)
  [arXiv:1702.00840 [nucl-ex]].
  
\bibitem{Chen:2016vem} 
  L.~Chen, G.~Y.~Qin, S.~Y.~Wei, B.~W.~Xiao and H.~Z.~Zhang,
  Phys.\ Lett.\ B {\bf 773}, 672 (2017)
  [arXiv:1607.01932 [hep-ph]].
  
\bibitem{Mueller:2016gko} 
  A.~H.~Mueller, B.~Wu, B.~W.~Xiao and F.~Yuan,
  Phys.\ Lett.\ B {\bf 763}, 208 (2016)
  [arXiv:1604.04250 [hep-ph]].
  
\bibitem{Lin:2014tya} 
  Z.~W.~Lin,
  Phys.\ Rev.\ C {\bf 90}, no. 1, 014904 (2014)
  [arXiv:1403.6321 [nucl-th]].

 
\end{thebibliography}
\end{document}